\documentclass{PoS}

\title{Integration-by-parts reductions from unitarity cuts and algebraic geometry}

\ShortTitle{Integration-by-parts reductions from unitarity cuts and algebraic geometry}

\author{\speaker{Kasper J. Larsen}\\
        Institute for Theoretical Physics, ETH Z{\"u}rich, 8093 Z{\"u}rich, Switzerland\\
        E-mail: \email{Kasper.Larsen@phys.ethz.ch}}

\author{Yang Zhang\\
        Institute for Theoretical Physics, ETH Z{\"u}rich, 8093 Z{\"u}rich, Switzerland\\
        E-mail: \email{Yang.Zhang@phys.ethz.ch}}

\abstract{Integration-by-parts reductions play a central role
in perturbative QFT calculations. They allow the set of Feynman
integrals contributing to a given observable to be reduced to a
small set of basis integrals, and they moreover facilitate the
computation of those basis integrals. We introduce an efficient
new method for generating integration-by-parts reductions. This
method simplifies the task by making use of generalized-unitarity
cuts and turns the problem of finding the needed total derivatives
into one of solving certain polynomial (so-called syzygy) equations.}

\FullConference{Loops and Legs in Quantum Field Theory\\
		24-29 April 2016\\
		Leipzig, Germany}

\usepackage{amsmath}
\usepackage{amssymb}
\usepackage{url}

\def\d{{\rm d}}

\begin{document}

\section{Introduction}

Precise predictions of the production cross sections
at the Large Hadron Collider are necessary to gain a
quantitative understanding of the Standard Model signals and background.
To match the experimental precision and the parton distribution
function uncertainties, this typically requires computations
at next-to-next-to leading order in fixed-order perturbation
theory. Calculations at this order are challenging
due to the large number of contributing Feynman diagrams,
involving loop integrals with high powers of loop momenta in the
numerator of the integrand.

A key tool in these computations are integration-by-parts (IBP)
identities \cite{Tkachov:1981wb,Chetyrkin:1981qh}. These are relations
that arise from the vanishing integration of total derivatives.
Schematically, they take the form,
\begin{equation}
\int \prod_{i=1}^L \frac{\d^D \ell_i}{\pi^{D/2}}
\sum_{j=1}^L \frac{\partial}{\partial \ell_j^\mu}
\frac{v_j^\mu \hspace{0.5mm} P}{D_1^{a_1} \cdots D_k^{a_k}}
\hspace{1mm}=\hspace{1mm} 0 \,, \label{eq:IBP_schematic}
\end{equation}
where $P$ and the vectors $v_j^\mu$ are polynomials in the internal and
external momenta, the $D_k$ denote inverse propagators, and
$a_i \geq 1$ are integers. In practice, the IBP identities
generate a large set of linear relations between loop integrals,
allowing a significant fraction to be reexpressed in terms of a finite
basis of integrals. (The fact that the basis of
integrals is always finite was proven in ref.~\cite{Smirnov:2010hn}.)
The latter step of solving the linear systems arising from
eq.~(\ref{eq:IBP_schematic}) may be done by Gauss-Jordan elimination
in the form of the Laporta algorithm~\cite{Laporta:2000dc,Laporta:2001dd},
leading in general to relations involving integrals with
squared propagators. There are several publically available implementations
of automated IBP reduction: AIR~\cite{Anastasiou:2004vj},
FIRE~\cite{Smirnov:2008iw,Smirnov:2014hma},
Reduze~\cite{Studerus:2009ye,vonManteuffel:2012np},
LiteRed~\cite{Lee:2012cn}, in addition to private implementations.
An approach for deriving IBP reductions without squared
propagators was developed in ref.~\cite{Gluza:2010ws}.
A recent approach \cite{vonManteuffel:2014ixa} uses numerical
sampling of finite-field elements to construct the reduction
coefficients.

In addition to reducing the contributing Feynman diagrams to
a small set of basis integrals, the IBP reductions
provide a way to compute these integrals themselves via
differential equations%
~\cite{Kotikov:1990kg,Kotikov:1991pm,Bern:1993kr,Remiddi:1997ny,Gehrmann:1999as,Henn:2013pwa,Ablinger:2015tua}.
Letting $x_m$ denote a kinematical variable,
$\epsilon = \frac{4-D}{2}$ the dimensional regulator, and
$\boldsymbol{\mathcal{I}}(\boldsymbol{x},\epsilon)
=\big(\mathcal{I}_1 (\boldsymbol{x},\epsilon),
\ldots, \mathcal{I}_N  (\boldsymbol{x},\epsilon) \big)$
the basis of integrals, the result of differentiating a basis integral
wrt. $x_m$ can be written as a linear combination
of the basis integrals by using, in practice, the IBP reductions.
As a result, one has a linear system of differential equations,
\begin{equation}
\frac{\partial}{\partial x_m} \boldsymbol{\mathcal{I}}(\boldsymbol{x},\epsilon)
= A_m (\boldsymbol{x}, \epsilon) \boldsymbol{\mathcal{I}}(\boldsymbol{x},\epsilon) \,,
\end{equation}
which, supplied with appropriate boundary conditions, can be
solved to yield expressions for the basis integrals.
This method has proven to be a powerful tool for compu\-ting two-
and higher-loop integrals. IBP reductions thus play a central role
in perturbative calculations in particle physics.

In many realistic multi-scale problems, such as $2 \to n$ scattering
amplitudes with $n\geq 2$, the step of generating IBP reductions with
existing algorithms is the most challenging part of the calculation.
It is therefore interesting to explore other approaches to generating
these reductions.

In these proceedings, based on ref.~\cite{Larsen:2015ped}, we
explain how IBP reductions can be obtained efficiently by applying
a set of unitarity cuts dictated by a specific set of subgraphs
and solving associated polynomial (syzygy) equations.
A similar approach was introduced by Harald Ita in ref.~\cite{Ita:2015tya}
where IBP relations are also studied in connection with cuts, and
their underlying geometric interpretation is clarified.

\section{Setup}\label{sec:Setup}

Throughout these proceedings, we will focus on the case of two-loop integrals.
We will denote the number of external legs of a given two-loop integral
by $n$, and the number of propagators by $k$. (Note that, after integrand reduction, $k \leq 11$.)
We work in dimensional regularization and use the
four-dimensional helicity scheme, taking the external momenta in four dimensions.

Our first aim is to recast two-loop integrals in a parametrization
which is useful for deriving IBP relations.
The first step is to decompose the loop momenta into four- and
$(D-4)$-dimensional parts, $\ell_i = \overline{\ell}_i + \ell_i^\perp$,
$i=1,2$. Next, parametrize the extra-dimensional vectors
$\ell_i^\perp$ in hyperspherical coordinates, using their norms
$\mu_{ii}\equiv-( \ell_i^\perp)^2 \geq 0$ and relative angle
$\mu_{12}\equiv -\ell_1^\perp \cdot \ell_2^\perp$. After this transformation,
the two-loop integral takes the form
\begin{align}
I^{(2)}_{n\geq 5} &= \frac{2^{D-6}}{\pi^{5}\Gamma(D-5) }\int_0^\infty \d \mu_{11} \int_0^\infty \d \mu_{22}
  \int_{-\sqrt{\mu_{11}\mu_{22}}}^{\sqrt{\mu_{11}\mu_{22}}} \d\mu_{12}
 \big(\mu_{11} \mu_{22}-\mu_{12}^2 \big)^{\frac{D-7}{2}}  \int \d^4
 \overline{\ell}_1  \hspace{0.5mm}  \d^4 \overline{\ell}_2 \frac{ P(\ell_1,\ell_2)}{D_1
   \cdots D_k} \,.
 \label{two-loop-integral}
\end{align}
For $n \geq 5$ there are $11-k$ irreducible scalar products (ISPs) which
we denote by $g_j$ where $j=1, \ldots, 11-k$. We can then define
variables $z_1,\ldots, z_{11}$ as follows,
\begin{equation}
z_i \equiv \left\{
\begin{array}{lrl}
  D_i     \hspace{5mm} & 1  \leq &\hspace{-1mm} i \leq k\\[1mm]
  g_{i-k} \hspace{5mm} & k+1\leq &\hspace{-1mm} i \leq m \,,
\end{array} \right.
\label{eq:2}
\end{equation}
with $m=11$ for $n\geq5$. The transformation $\{\bar \ell_i, \mu_{ij}\} \rightarrow \{z_1, \ldots, z_{11}\}$
is invertible, with a polynomial inverse (provided the
$g_j$ are chosen to take the form
$\frac{1}{2} (\ell_i + K_j)^2$, rather than linear
dot products of the $\ell_i$), and has a constant Jacobian.
The integral (\ref{two-loop-integral}) then becomes,
\begin{equation}
I^{(2)}_{n\geq 5} = \frac{2^{D-6}}{\pi^{5}\Gamma(D-5) J} \int \prod_{i=1}^{11} \d z_i \hspace{0.6mm}
  F(z)^{\frac{D-7}{2}} \frac{P(z)}{z_1 \cdots z_k}\,,
\label{two-loop-integral-z}
\end{equation}
where $F(z)$ denotes the kernel $(\mu_{11} \mu_{22}-\mu_{12}^2)$
expressed in the $z_i$, in which it is polynomial.

The representation (\ref{two-loop-integral-z}) is valid for $n \geq 5$ external legs. For
lower multiplicities, the loop momenta have components
which can be integrated out before the transformation
(\ref{eq:2}) is applied. For example, for $n=4$, by momentum
conservation, there are only three linearly independent external
momenta, and we can find an orthogonal vector $\omega$; that is,
$p_i \cdot \omega=0$ where $i=1,\ldots, 4$. Integrating out the components $\ell_i \cdot \omega$
in eq.~(\ref{two-loop-integral}) and subsequently
applying the transformation (\ref{eq:2}) yields,
\begin{equation}
I^{(2)}_{n=4} = \frac{2^{D-5}}{\pi^{4}\Gamma(D-4) J} \int \prod_{i=1}^{9} \d z_i \hspace{0.6mm}
  F(z)^{\frac{D-6}{2}} \frac{P(z)}{z_1 \cdots z_{k}} \,.
\label{two-loop-integral-4z}
\end{equation}
We observe that the spacetime dimensions appearing here
have been shifted down by one relative to eq.~(\ref{two-loop-integral-z});
that is, $D - D_c \to D - (D_c - 1)$. This is consistent
with the fact that the span of the external momenta has one dimension less.
For notational convenience we will drop the prefactors
in front of the integral signs in eqs.~(\ref{two-loop-integral-z})
and (\ref{two-loop-integral-4z}). We note that the representations
(\ref{two-loop-integral-z}) and (\ref{two-loop-integral-4z})
have also been considered in the literature by Baikov, see for example
ref.~\cite{Baikov:1996rk}.

\section{Integration-by-parts reductions on generalized-unitarity cuts}

The key idea of our approach is to derive IBP identities on
\emph{generalized-unitarity cuts} where some subset $S$ of
propagators are put on shell: $D_i^{-1} \to \delta(D_i)$ with
$i \in S$. In essence, the application of cuts divide the task of
finding IBP reductions into several smaller, and more manageable,
problems. This is because, on a given cut $S$, only integrals
which contain all of the propagators in $S$ contribute (since
the remaining integrals are missing the pole whose residue the
cut is computing). As the resulting IBP identities will miss
contributions from some of the basis integrals, we must construct
the identities on a \emph{set of cuts} $S_1, \ldots, S_C$
and merge the partial results. Below we explain how to construct
IBP identities on a given cut $S$ and how to choose an appropriate
spanning set of cuts.

Let us consider a cut where $c$ propagators are put on shell ($0\leq c\leq k$).
We label the propagators of the graph (cf.~the labelling, e.g., in Fig.~\ref{fig:dbox_cut})
and let $\mathcal{S}_\mathrm{cut}$, $\mathcal{S}_\mathrm{uncut}$ and
$\mathcal{S}_\mathrm{ISP}$ denote the sets of indices labelling
cut propagators, uncut propagators and ISPs, respectively.
$\mathcal{S}_\mathrm{cut}$ thus contains $c$ elements. Furthermore, we let $m$
denote the total number of $z_j$ variables, and
set $\mathcal{S}_\mathrm{uncut}=\{r_1,\ldots, r_{k-c}\}$
and $\mathcal{S}_\mathrm{ISP}=\{r_{k-c+1},\ldots, r_{m-c}\}$.
Then, after cutting the propagators, $z_i^{-1} \to \delta(z_i),
i \in \mathcal{S}_\mathrm{cut}$, the integrals (\ref{two-loop-integral-z}) and
(\ref{two-loop-integral-4z}) reduce to,
\begin{equation}
I^{(2)}_\mathrm{cut} = \int \frac{\d z_{r_1} \cdots \d z_{r_{m-c}} P(z)}
{z_{r_1}\cdots z_{r_{k-c}}} F(z)^{\frac{D - h}{2}} \bigg|_{z_i=0\,, \forall i\in \mathcal{S}_\mathrm{cut}} \,,
\label{cut-z}
\end{equation}
where $h$ depends on the number
of external legs: $h=6$ for $n=4$ and $h=7$ for
$n \geq 5$.

Now we turn to the problem of writing down IBP relations.
An IBP relation (\ref{eq:IBP_schematic})
that concerns integrals with $m$ integration variables
corresponds to a total derivative, or equivalently
an exact differential form, of degree $m$. Here we are interested
in the $c$-fold cut of the IBP relation,
where the propagators of $\mathcal{S}_\mathrm{cut}$ are
put on shell in all terms (and integrals which do not contain all of these propagators
are set to zero). Such $c$-fold cut relations
correspond to exact differential forms of degree $m-c$.
The generic exact form that matches
the form of the integrand in eq.~(\ref{cut-z}) is,
\begin{equation}
  \label{ansatz}
0=\int \d \bigg( \sum_{i=1}^{m-c} \hspace{-0.8mm}
     \frac{(-1)^{i+1} a_{r_i} F(z)^{\frac{D-h}{2}}}{z_{r_1}\cdots z_{r_{k-c}}}
     \d z_{r_1} \hspace{-0.5mm} \wedge \cdots \wedge \widehat{\d z_{r_i}} \wedge \cdots
     \wedge \hspace{-0.5mm} \d z_{r_{m-c}} \hspace{-0.5mm} \bigg)
\end{equation}
{\vskip -1.5mm}
\noindent where the $a_i$'s are polynomials in $\{z_{r_1}, \ldots,
z_{r_{m-c}}\}$. (Similar differential form ans{\"a}tze for four-dimensional
IBPs on cuts were considered in ref.~\cite{Zhang:2014xwa}.)
Writing out eq.~(\ref{ansatz}) more explictly, we get the IBP relation,
\begin{equation}
  0=\int \bigg(\sum_{i=1}^{m-c} \Big(\frac{\partial a_{r_i} }{\partial z_{r_i}}
  + \frac{D-h}{2F}a_{r_i}\frac{\partial F }{\partial
    z_{r_i}}\Big)-\sum_{i=1}^{k-c} \frac{a_{r_i}}{z_{r_i}}\bigg)
    \frac{F(z)^{\frac{D-h}{2}}}{z_{r_1}\cdots z_{r_{k-c}}} \d z_{r_1} \wedge \cdots
  \wedge \d z_{r_{m-c}} \,. \label{IBP}
\end{equation}
Now, for generic $a_i$, the second term in the sum corresponds to an integrand in
$(D-2)$ dimensions. This is because the factor $F$ in the denominator
divides against the integration measure $F(z)^\frac{D-h}{2}$ and thereby
shifts $D \to D-2$ in the exponent. Likewise, the third term
generates integrals with doubled propagators. To get an IBP relation that
involves only integrals in $D$ dimensions with single-power
propagators, we require the $a_i$ to be such that the second
and third terms of (\ref{IBP}) are \emph{polynomial} rather than
rational,
\begin{align}
b F + \sum_{i=1}^{m-c} a_{r_i}\frac{\partial F}{\partial
z_{r_i}} &=0 \label{eq:syzygy_1} \\
a_{r_i} + b_{r_i} z_{r_i}&=0\,, \hspace{4mm} i=1,\ldots, k-c \,,
\label{eq:syzygy_2}
\end{align}
where $a_{r_i}$, $b$ and $b_{r_i}$ must be polynomials in $z_j$.
Equations of this type are known in algebraic geometry as
\emph{syzygy equations}. They have been considered in the
context of IBP relations in refs.~\cite{Gluza:2010ws,Schabinger:2011dz,Ita:2015tya}.
In practice, the equations (\ref{eq:syzygy_1})--(\ref{eq:syzygy_2})
can be solved with computational algebraic geometry software, such as Singular~\cite{DGPS}.
The corresponding IBP identities are then obtained by
plugging the solutions into eq.~(\ref{IBP}).
In this way, we find IBP identities on the cut
$\mathcal{S}_\mathrm{cut}$.

To find an appropriate set of cuts on which to reconstruct the
IBP reductions, we first find a basis of integrals. We perform
this step without applying cuts and using rational numbers or finite-field element
values for the kinematical invariants and spacetime dimension. Applying
Gauss-Jordan elimination to the resulting set of IBP identities with some chosen
ordering on the set of integrals then produces a basis of integrals.
(Similar ideas for finding a basis of integrals have
appeared in ref.~\cite{RobRadCor2013}, using random prime numbers for the
external invariants and spacetime dimension, and in ref.~\cite{Kant:2013vta},
using finite-field elements.)
In the example in the next section we explain how the set of cuts
is obtained from the basis of integrals. Having obtained
an appropriate set of cuts, we proceed analytically
and construct IBP reductions on each cut in turn. Finally, we
merge the results obtained from the cuts to find complete IBP reductions.

\section{Example}

\begin{figure}[!h]
\includegraphics[scale=0.8]{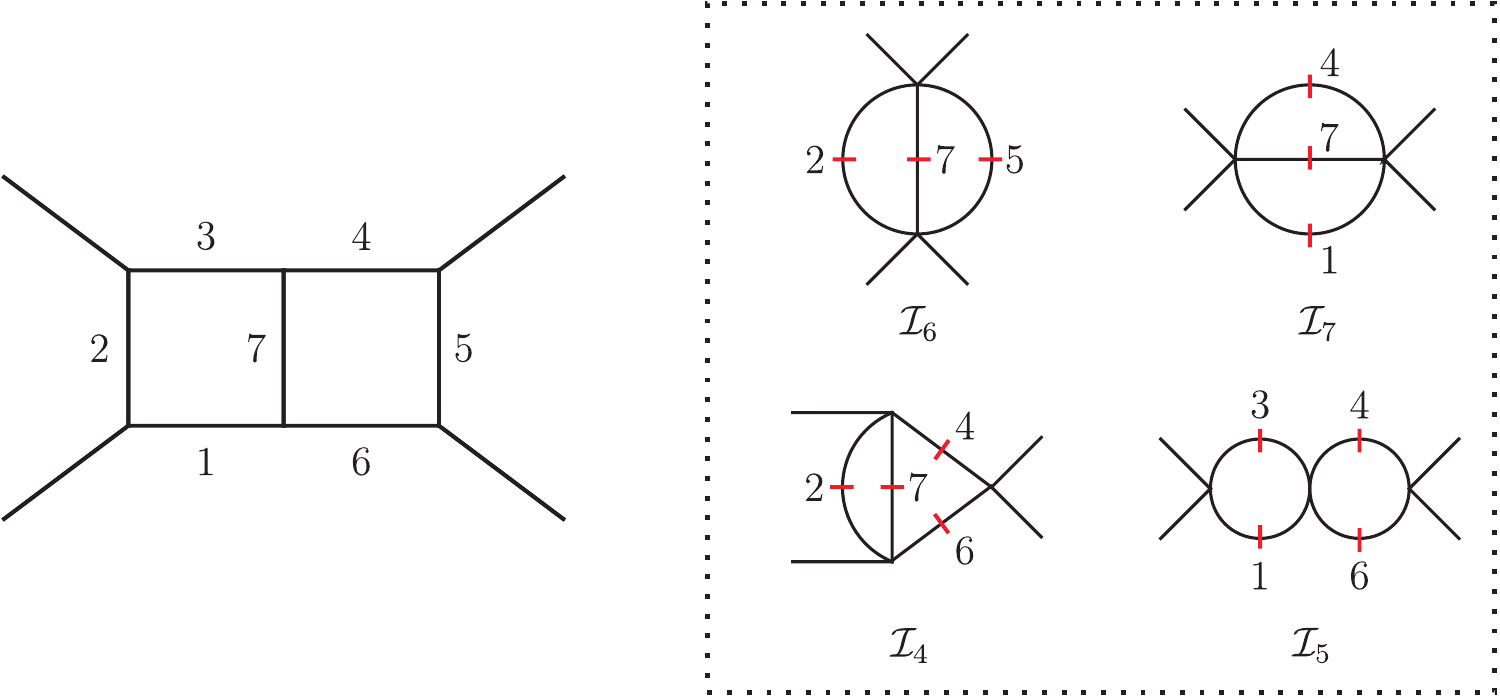}
\caption{The massless double-box diagram, along with our labelling
 conventions for its internal lines, is shown on the left. The right
part shows the subset of the master integrals in eq.~(\protect\ref{eq:master_integrals})
 with the property that their graphs cannot be obtained by
 adding internal lines to the graph of some other master integral.
The corresponding cuts $\{2,5,7\}$, $\{1,4,7\}$, $\{2,4,6,7\}$ and $\{1,3,4,6\}$
are the cuts required for deriving complete
 IBP relations of integrals with this double-box topology.}
\label{fig:dbox_cut}
\end{figure}

To demonstrate the method, we consider the example of a planar double-box integral
with all legs and propagators massless, illustrated in Fig.~\ref{fig:dbox_cut}.
For this integral we have $k=7$, and the inverse propagators can be parametrized as,
\begin{align}
  D_1 &=\ell_1^2 \,,\quad D_2 =(\ell_1-p_1)^2\,,\quad D_3
  =(\ell_1-p_1-p_2)^2 \nonumber \\
D_4 &=(\ell_2-p_3-p_4)^2\,,\quad D_5 =(\ell_2-p_4)^2\,,\quad D_6
  =\ell_2^2\,,\quad D_7=(\ell_1+\ell_2)^2 \,. \label{eq:3}
\end{align}
As mentioned above eq.~(\ref{eq:2}), the generic integral
of this topology will have numerator insertions which are
monomials in two distinct ISPs. The ISPs may be chosen as,
\begin{align}
D_8=\frac{1}{2}(\ell_1+p_4)^2,\quad D_9=\frac{1}{2}(\ell_2+p_1)^2 \,.
\label{eq:5}
\end{align}
Our aim is now to show how the IBP reductions of an
integral with the propagators in eq.~(\ref{eq:3}) with a generic numerator
insertion can be obtained. After the change of variables
$\{\bar \ell_i, \mu_{ij}\} \rightarrow \{z_1, \ldots, z_9\}$ discussed
in section~\ref{sec:Setup}, the double-box integral takes the form
of eq.~(\ref{two-loop-integral-4z}).

We will use the following notation for the integrals,
\begin{equation}
  \label{eq:4}
  G[n_1,\ldots,n_9] \equiv \int \prod_{i=1}^{9} \d z_i \hspace{0.6mm}
  F(z)^{\frac{D-6}{2}} z_1^{n_1} \cdots z_9^{n_9} \,.
\end{equation}

Now, to find the IBP reductions of these integrals,
the first step is to find a basis of integrals. This is done by solving the
syzygy equations (\ref{eq:syzygy_1})--(\ref{eq:syzygy_2})
without imposing cuts while using numerical external kinematics
(with rational numbers or finite-field elements), and then inserting
all solutions into the right-hand side of eq.~(\ref{IBP}),
and finally performing Gauss-Jordan elimination with some
chosen ordering on the set of integrals.
In the case at hand, we find the following set of master integrals
(after modding out by symmetries)
\begin{equation}
\begin{array}{rl}
\mathcal{I}_1 \hspace{0mm} &\equiv \hspace{0mm} G[-1,-1,-1,-1,-1,-1,-1,0,0]\\
\mathcal{I}_2 \hspace{0mm} &\equiv \hspace{0mm} G[-1,-1,-1,-1,-1,-1,-1,1,0]\\
\mathcal{I}_3 \hspace{0mm} &\equiv \hspace{0mm} G[0, -1, -1, 0, -1, -1, -1, 0, 0]\\
\mathcal{I}_4 \hspace{0mm} &\equiv \hspace{0mm} G[0, -1, 0, -1, 0, -1, -1, 0, 0] \\
\mathcal{I}_5 \hspace{0mm} &\equiv \hspace{0mm} G[-1, 0, -1, -1, 0, -1, 0, 0, 0] \\
\mathcal{I}_6 \hspace{0mm} &\equiv \hspace{0mm} G[0, -1, 0, 0, -1, 0, -1, 0, 0]\\
\mathcal{I}_7 \hspace{0mm} &\equiv \hspace{0mm} G[-1, 0, 0, -1, 0, 0, -1, 0, 0] \\
\mathcal{I}_8 \hspace{0mm} &\equiv \hspace{0mm} G[-1, -1, -1, 0, -1, 0, -1, 0, 0] \,.
\end{array}
\label{eq:master_integrals}
\end{equation}
Having obtained a basis of integrals, we proceed to find
the IBP reductions analytically on a set of cuts and then merging
the results to find the complete reductions. To decide on the
minimal set of cuts required, we select those basis integrals
with the property that their graphs cannot be obtained by
adding internal lines to the graph of some integral in the basis.
In the present case, this subset is $\{ \mathcal{I}_4, \mathcal{I}_5,
\mathcal{I}_6, \mathcal{I}_7\}$, shown in Fig.~\ref{fig:dbox_cut}.
Hence, we only need to consider the four cuts $\{2,5,7\}$, $\{1,4,7\}$,
$\{2,4,6,7\}$ and $\{1,3,4,6\}$ to find the complete IBP reductions.

To illustrate how to find the IBP reductions on a given cut,
let us consider the three-fold cut $\mathcal{S}_\mathrm{cut}=\{2,5,7\}$.
Here, $\mathcal{S}_\mathrm{uncut}=\{1,3,4,6\}$ and $\mathcal{S}_\mathrm{ISP}=\{8,9\}$. The
kernel $F$ on the cut is
polynomial in $z_1$, $z_3$, $z_4$, $z_6$, $z_8$ and $z_9$. The syzygy equations
(\ref{eq:syzygy_1})--(\ref{eq:syzygy_2}) read,
\begin{equation}
b F \hspace{0.5mm} + \sum_{i\in\{ 1,3,4,6,8,9\}} a_i \frac{\partial F}{\partial
z_i} =0 \hspace{9mm} \mathrm{and} \hspace{9mm} a_j + b_j z_j =0\,, \hspace{3mm} j\in \{1,3,4,6\} \,,
\label{dbox_sunset1_syz}
\end{equation}
where $b, b_i, a_j$ are to be solved for as polynomials in $z_k$.
A generating set of solutions of eq.~(\ref{dbox_sunset1_syz}) can
be found via algebraic geometry software such as Singular
in seconds (with analytic coefficients). Now, given
a solution $(b, b_i, a_j)$, any multiple
$(q b, q b_i, q a_j)$, with $q$ a polynomial,
is manifestly also a solution. To capture the IBP reductions of
all possi\-ble numerator insertions, we thus consider all syzygies
$(b, b_i, a_j)$ multiplied by appropriate
monomials in the ISPs, $q = \prod_{i \in \{ 1,3,4,6,8,9 \}} D_i^{a_i}$.
Inserting all such solutions into the right-hand side
of eq.~(\ref{IBP}) produces the complete set of
IBP relations, without doubled
propagators, on the cut $\{2,5,7\}$ (i.e., up to integrals that vanish on this cut).

As an example, consider the tensor integral $T\equiv G[-1,-1,-1,-1,-1,-1,-1,0,2]$.
On the four cuts specified above this integral reduces to, respectively,
\begin{align}
T\big|_{\{2,5,7\}}   &= \hspace{0mm} \sum_{j\in \{ 1,2,3,6,8 \}} \hspace{-2mm}
c_j \mathcal{I}_j \,, \hspace{6mm} T\big|_{\{1,4,7\}}    = \hspace{0mm}
\sum_{j\in \{ 1,2,7 \}} \hspace{-1mm} c_j \mathcal{I}_j \,,
\label{eq:IBP_on_cut_formal_1}\\
T\big|_{\{2,4,6,7\}} &= \hspace{0mm} \sum_{j\in \{ 1,2,4 \}} \hspace{-1mm}
c_j \mathcal{I}_j \,, \hspace{6mm} T\big|_{\{1,3,4,6\}}  = \hspace{0mm}
\sum_{j\in \{ 1,2,5 \}} \hspace{-1mm} c_j \mathcal{I}_j \,,
\label{eq:IBP_on_cut_formal_2}
\end{align}
where, denoting $\chi \equiv t/s$, the coefficients are found to be,
\begin{align}
c_1 &= \frac{(D-4) s^2\chi}{8 (D-3)}\,, \hspace{8mm}
c_2 = -\frac{(3 D-2 \chi -12)s}{4 (D-3)} \,,\hspace{8mm}
c_3 = \frac{(4-D) (9 \chi +7)}{4 (D-3)} \\
c_4 &= \frac{(10 -3D) (2 \chi -13)}{8 (D-4)s} \,, \hspace{8mm}
c_5 = \frac{2 D (\chi +1){-}8 \chi{-}7}{2 (D-4)s} \,, \hspace{8mm}
c_6 = \frac{9 (3D{-}10) (3D{-}8)}{4 (D-4)^2 s^2\chi} \\
c_7 &= \frac{(3 D-10) (3 D-8) (2 \chi +1)}{2 (D-4)^2 (D-3) s^2} \,, \hspace{8mm}
c_8 = 2 \,.
\end{align}
The integrals absent from the right-hand sides of
eqs.~(\ref{eq:IBP_on_cut_formal_1})--(\ref{eq:IBP_on_cut_formal_2})
vanish on the respective cuts. Combining these results,
we get the \emph{complete} IBP reduction of the tensor integral,
\begin{equation}
T = \sum_{j=1}^8 c_j \mathcal{I}_j \,.
\label{eq:complete_IBP_of_tensor}
\end{equation}
We have implemented the algorithm as a program, powered by Mathematica
and Singular~\cite{DGPS}. It analytically reduces all integrals with numerator
rank $\leq 4$, to the eight master integrals in
eq.~(\ref{eq:master_integrals}) in about $39$ seconds in the fully
massless case, and to $19$ master integrals in about $162$ seconds in the
one-massive-particle case (on a laptop with 2.5 GHz Intel Core i7 and 16 GB RAM).

One important feature of the approach is the use of the $z_i$-variables
in eq.~(\ref{eq:2}) which
ultimately lead to the simple form of the syzygy equations (\ref{eq:syzygy_1})--(\ref{eq:syzygy_2}).
However, the crucial feature is the use of generalized-unitarity cuts: they eliminate
variables in the syzygy equations so that these can be solved
more efficiently. More importantly, because on any given cut, only
a subset of basis integrals contribute, the cuts have the effect of
``block-diagonalizing'' the linear system of IBP identities on which
Gauss-Jordan elimination is performed to find the IBP reductions.

There are several directions
for future research. Of particular interest are extensions to
higher multiplicity, several external and internal masses,
non-planar diagrams, and higher loops.

{\bf Acknowledgments}

We thank S.~Badger, H.~Frellesvig, E.~Gardi, A.~Georgoudis, A.~Huss, H.~Ita,
D.~Kosower, A.~von~Manteuffel, M.~Martins, C.~Papadopoulos and R.~Schabinger for useful discussions.
The research leading to these results has received
funding from the European Union Seventh Framework
Programme (FP7/2007-2013) under grant agreement no.~627521, and Swiss
National Science Foundation (Ambizione~PZ00P2\_161341).

\end{document}